\documentclass[preprint]{aastex}
\begin{document}

\shortauthors{Welty \& Fitzpatrick}
\shorttitle{Variable IS Absorption}

\title  {Variable Interstellar Absorption toward the Halo Star HD 219188 --- Implications for Small-Scale Interstellar Structure
\footnote{Based in part on observations with the NASA/ESA {\it Hubble Space Telescope}, obtained from the data Archive at the Space Telescope Science Institute, which is operated by the Association of Universities for Research in Astronomy, Inc., under NASA contract NAS5-26555.}}

\author{Daniel E. Welty\altaffilmark{2,3,4}}
\affil{University of Chicago, Astronomy and Astrophysics Center, 5640 S. Ellis Ave., Chicago, IL  60637}
\email{welty@oddjob.uchicago.edu}
\altaffiltext{2}{Visiting observer, Kitt Peak National Observatory, National Optical Astronomy Observatories, which is operated by the Association of Universities for Research in Astronomy, Inc. (AURA) under cooperative agreement with the National Science Foundation}
\altaffiltext{3}{Visiting observer, Anglo-Australian Observatory}
\altaffiltext{4}{Visiting observer, European Southern Observatory (Chile); program 64.I-0475}

\author{Edward L. Fitzpatrick\altaffilmark{2}}
\affil{Villanova University, Department of Astronomy \& Astrophysics, 800 Lancaster Ave., Villanova, PA  19085}
\email{fitz@ast.vill.edu}

\begin{abstract}

Within the last 10 years, strong, narrow \ion{Na}{1} absorption has appeared 
at $v_{\sun}$ $\sim$ $-$38 km~s$^{-1}$ toward the halo star HD 219188;
that absorption has continued to strengthen, by a factor 2--3, over the 
past three years.
The line of sight appears to be moving into/through a relatively cold, 
quiescent intermediate velocity (IV) cloud, due to the 13 mas/yr proper 
motion of HD 219188; the variations in \ion{Na}{1} probe length scales of 2--38 AU/yr.
UV spectra obtained with the {\it HST} GHRS in 1994--1995 suggest $N$(H$_{\rm tot}$) $\sim$ 4.8 $\times$ 10$^{17}$ cm$^{-2}$, ``halo cloud'' depletions, $n_{\rm H}$ $\sim$ 25 cm$^{-3}$, and $n_e$ $\sim$ 0.85--6.2 cm$^{-3}$ (if $T$ $\sim$ 100 K) for the portion of the IV cloud sampled at that time.
The relatively high fractional ionization, $n_e$/$n_{\rm H}$ $\ga$ 0.034, implies that hydrogen must be partially ionized. 
The $N$(\ion{Na}{1})/$N$(H$_{\rm tot}$) ratio is very high; in this case, the variations in \ion{Na}{1} do not imply large local pressures or densities.

\end{abstract}


\section{Introduction}
\label{sec-intro}

Interstellar absorption lines were first recognized as interstellar by virtue 
of their invariance amidst the shifting stellar lines in the $\delta$ Ori A 
spectroscopic binary system (Hartmann 1904).
Recently, however, it has become evident that some clearly interstellar 
lines do vary in strength and/or velocity.
Variations in \ion{Na}{1} and/or \ion{Ca}{2} have been observed toward several 
stars in the direction of the Vela supernova remnant (Hobbs et al. 1991; 
Danks \& Sembach 1995; Cha \& Sembach 2000) --- presumably due to cloud 
motions in that energetic environment.
High $N$(\ion{Ca}{2})/$N$(\ion{Na}{1}) ratios suggest significant shock 
processing of the interstellar dust grains in some of those clouds.
In several other cases, however, the narrow \ion{Na}{1} line widths 
($b$ $\sim$ 0.3--0.5 km s$^{-1}$) and/or low $N$(\ion{Ca}{2})/$N$(\ion{Na}{1}) 
ratios characterizing the variable components suggest that the gas is relatively
cool and quiescent (Blades et al. 1997; Price 
et al. 2000; Crawford et al. 2000; Lauroesch, Meyer, \& Blades 2000).
Detailed information regarding the physical conditions in these varying components is generally not available, however.

These indications of temporal variability may be related to recent
results on small scale spatial structure in the ISM.
Observations of \ion{H}{1} 21cm absorption toward high-velocity pulsars have
revealed structure on scales of 5--100 AU (Frail et al. 1994). 
Observations of absorption due to the trace ions \ion{Na}{1} and/or \ion{K}{1} toward globular clusters have found significant differences  on scales of 10$^3$--10$^6$ AU (Langer, Prosser, \& Sneden 1990; Kemp, Bates, \& Lyons 1993; Meyer \& Lauroesch 1999).
Differences in \ion{Na}{1} and/or \ion{K}{1} absorption toward the individual 
members of binary or multiple stellar systems are commonplace, suggesting
ubiquitous structure on intermediate scales of 10$^2$--10$^4$ AU
(Watson \& Meyer 1996; Lauroesch \& Meyer 1999).
If these differences are due to variations in overall hydrogen column density
[assuming ``typical'' $N$(\ion{Na}{1})/$N$(H) ratios], the high implied 
densities (generally several thousand cm$^{-3}$) are difficult to 
reconcile with clouds in thermal pressure equilibrium at typical interstellar 
pressures of order 3000 cm$^{-3}$K.
The differences may reflect a population of cold, dense filaments or sheets, 
embedded in warmer, less dense neutral gas and containing 10--30\% of the total 
column density of cold, neutral gas (Heiles 1997).
Alternatively, the differences in the trace ions might be due to variations in density
and/or ionization on those small scales (e.g., Lauroesch et al. 1998).
Differences in ionization would not explain the variations seen in \ion{H}{1}, however.

HD 219188 is an early B supergiant located about 2800 pc from the sun at 
($l$,$b$) $\sim$ (83\arcdeg,$-$50\arcdeg).
Because of its location and brightness ($V$ $\sim$ 6.9), it has been 
observed in a number of studies investigating the interstellar clouds 
in the lower Galactic halo (M\"{u}nch \& Zirin 1961; Albert 1983; Danly 1989; 
Sembach, Danks, \& Savage 1993).
Strong, multi-component absorption from \ion{Na}{1} and \ion{Ca}{2} is apparent 
for $-$17 km~s$^{-1}$ $\la$ $v_{\sun}$ $\la$ +5 km~s$^{-1}$; weaker 
\ion{Ca}{2} absorption extends to about $-$35 km~s$^{-1}$.
High resolution \ion{Na}{1} spectra obtained in 1997,
however, revealed a new, relatively strong, narrow 
component at $v$ $\sim$ $-$38 km~s$^{-1}$.
This component is also present in \ion{Ca}{2} and in several species in UV
(GHRS echelle) spectra obtained in 1994--1995.
Subsequent monitoring of the line of sight shows that the column 
densities of both \ion{Na}{1} and \ion{Ca}{2} have increased 
by additional factors of 2--3 since 1997.
In the following sections, we discuss the recent optical and UV spectra of HD 219188 and the inferred properties of the newly-revealed intermediate-velocity cloud along that line of sight.

\section{Optical Data}
\label{sec-optical}

The initial \ion{Na}{1} and \ion{Ca}{2} spectra of HD 219188 were obtained with the coud\'{e} feed and coud\'{e} spectrograph at the Kitt Peak National Observatory, as part of a program to acquire high resolution (FWHM $\la$ 1.5 km s$^{-1}$) optical spectra of halo stars previously observed at slightly lower resolution in the UV with the {\it HST} GHRS (see Fitzpatrick \& Spitzer 1997).
After the 1997 \ion{Na}{1} spectrum of HD 219188 revealed the new intermediate-velocity (IV) component at $-$38 km s$^{-1}$, additional spectra were obtained during various runs at Kitt Peak, the Anglo-Australian Observatory, and the European Southern Observatory in order to monitor possible further variations (Table~\ref{tab:opt}).

\begin{figure}
\epsscale{.9}
\plottwo{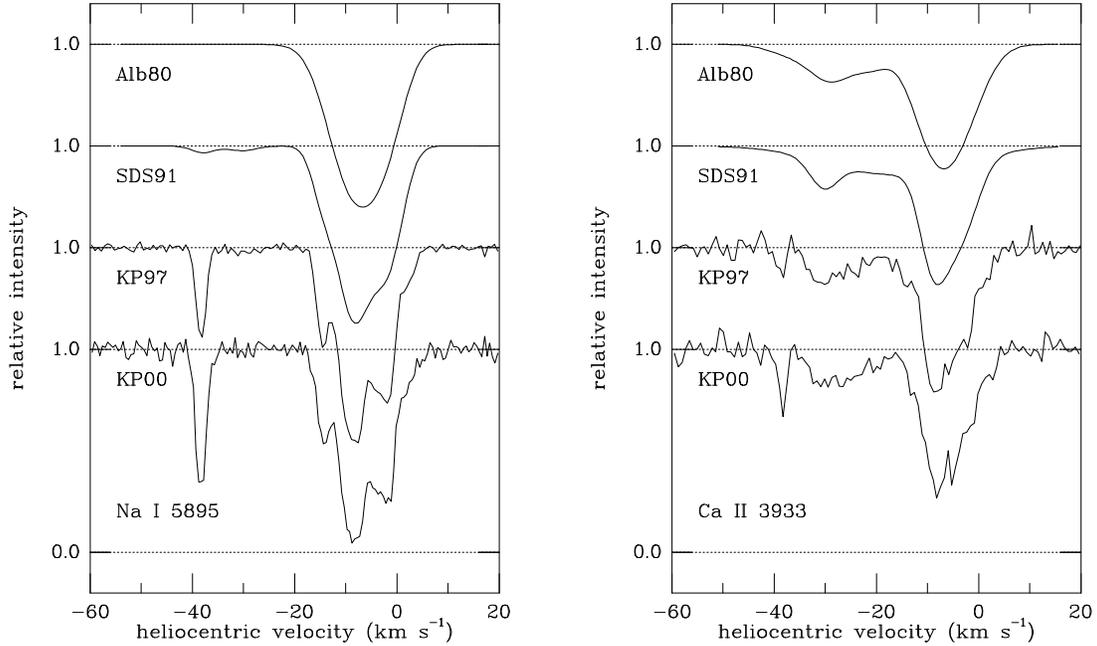}{h219v2ca.eps}
\caption{Selected Na I $\lambda$5895 (left) and Ca II $\lambda$3933 (right) spectra toward HD 219188.
The sources and dates of the spectra are indicated; see Table 1.
The column densities in the intermediate-velocity component at
$v$ $\sim$ $-$38 km s$^{-1}$ have increased by factors of 2--3
over the past 3--5 years; no Na I was detected at that velocity in 1980.
The lower resolution spectra from Albert (1983; Alb80) and Sembach et al. (1993; SDS91) were generated from the component structures listed therein.}
\label{fig:opt}
\end{figure}

The spectra were processed and extracted using standard procedures within IRAF, then normalized via polynomial fits to continuum regions adjacent to the absorption lines.
The resulting line profiles were fitted with multiple Voigt profile components, yielding column densities, line widths, and heliocentric velocities for the individual components (e.g., Welty, Morton, \& Hobbs 1996).
Some of the \ion{Na}{1} and \ion{Ca}{2} profiles are shown in Figure~\ref{fig:opt}, with the earlier profiles described by Albert (1983) and Sembach et al. (1993) at the top for comparison.

The changes in the strength of the $-$38 km s$^{-1}$ component --- over the past 20 years and over intervals as short as 6 months --- are quite evident and striking (Fig.~\ref{fig:opt}; Table~\ref{tab:opt}).
While weak \ion{Na}{1} absorption at $v$ $\sim$ $-$38 km s$^{-1}$ appears to have been present in 1991 (Sembach et al. 1993), none was detected in 1980 (Albert 1983). 
By 1997, the \ion{Na}{1} column density was roughly a factor of 10 higher than the limit in Albert (1983), and it has increased by a further factor of 2 since then.
As corresponding IV \ion{Ca}{2} absorption may be present in the profiles from both earlier studies (Fig.~\ref{fig:opt}), we have fitted those profiles using the component velocity structure derived from our higher resolution spectra to estimate $N$(\ion{Ca}{2}) in the IV cloud at those earlier epochs.
While our \ion{Ca}{2} column densities in 1995 and 1997 may be consistent with the (uncertain) value reported by Albert (1983), $N$(\ion{Ca}{2}) has increased by a factor of 2--3 since then.
Over the past 3--5 years, the $N$(\ion{Na}{1})/$N$(\ion{Ca}{2}) ratio thus 
appears to have remained roughly constant (though significantly 
higher than previous values or limits) (Table~\ref{tab:opt}).

In contrast to the changes in column density, neither the velocity of the IV component nor the \ion{Na}{1} line width has changed significantly since 1995.
The width of the \ion{Na}{1} line, $b$ $\sim$ 0.55--0.60 km s$^{-1}$ implies a maximum temperature for the cloud of 490 K and a maximum 3-dimensional internal 
turbulent velocity (3$^{1/2}$)$v_t$ = 0.73 km~s$^{-1}$ (comparable to the 
sound speed for $T$ $\sim$ 80--100 K).

\section{UV Data}
\label{sec-uv}

HD 219188 was observed with the {\it HST} GHRS on 1994 June 6 and 1995 May 16--17, at resolutions of about 3.5 km s$^{-1}$, under GTO program 5093.
Twelve wavelength settings were obtained, using the FP-SPLIT procedure to enable the identification and elimination of detector fixed-pattern features.
The individual sub-exposures for each wavelength setting were combined using the POFFSETS and SPECALIGN routines within the STSDAS package.
The summed spectra were normalized via polynomial fits to continuum regions adjacent to the absorption lines; the empirical S/Ns range from 30--50.
The interstellar absorption line profiles were then fitted using the component structures obtained from the higher resolution spectra of \ion{Na}{1} and \ion{Ca}{2} (Fitzpatrick \& Spitzer 1997; Welty et al. 1999).

\begin{figure}
\epsscale{.4}
\plotone{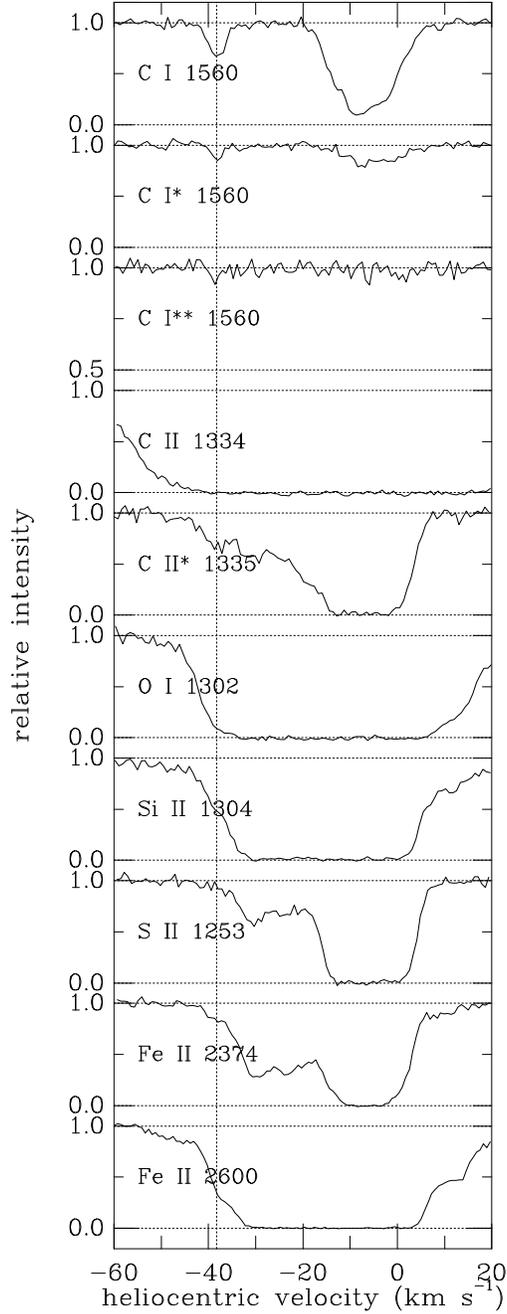}
\caption{Profiles of selected UV lines toward HD 219188, observed with GHRS at resolutions of about 3.5 km s$^{-1}$ in 1994--1995.
The vertical dotted line indicates the component at $-$38 km s$^{-1}$.}
\label{fig:ghrs}
\end{figure}

The $-38$ km s$^{-1}$ component is detected in the ground state lines of \ion{C}{1}, \ion{C}{2}, \ion{O}{1}, \ion{Si}{2}, \ion{S}{2}, and \ion{Fe}{2} and in the excited fine-structure lines of \ion{C}{1}*, \ion{C}{1}**, and \ion{C}{2}* (Table~\ref{tab:uv}, Fig.~\ref{fig:ghrs}).
If sulfur is undepleted, then $N$(\ion{S}{2}) suggests a total IV hydrogen column density $N$(H$_{\rm tot}$) = $N$(\ion{H}{1}) + $N$(\ion{H}{2}) $\sim$ 4.8 $\times$ 10$^{17}$ cm$^{-2}$.
The IV column density estimated for \ion{O}{1} (which should track \ion{H}{1}) is consistent with that $N$(H$_{\rm tot}$) and the typical interstellar oxygen abundance [O/H] $\sim$ $-$0.4 dex (Meyer, Jura, \& Cardelli 1998) --- which suggests that the IV gas is largely neutral.
The \ion{C}{2} $\lambda$1334 line is saturated, but we may use $N$(\ion{S}{2}) to estimate the total IV carbon abundance [$N$(\ion{C}{1}) + $N$(\ion{C}{2})] $\sim$ 6.8$\pm$2.1 $\times$ 10$^{13}$ cm$^{-2}$, as [C/S] $\sim$ $-$0.4$\pm$0.1 dex in diffuse Galactic clouds.
The column densities for \ion{Fe}{2} and \ion{Si}{2} yield depletions similar to those found for clouds in the Galactic halo (Welty et al. 1999, and references therein).

Analysis of the \ion{C}{1} fine-structure excitation equilibrium enables estimates for the thermal pressure and local density in the IV gas (Jenkins \& Shaya 1979).
The observed $N$(\ion{C}{1}*)/$N$(\ion{C}{1}) = 0.17$\pm$0.03 and $N$(\ion{C}{1}**)/$N$(\ion{C}{1}) = 0.06$\pm$0.02 yield $n_{\rm H}T$ $\sim$ 2500 cm$^{-3}$K and $n_{\rm H}$ $\sim$ 25 cm$^{-3}$ for $T$ = 100 K.
The thickness of the region probed, $N$(H)/$n_{\rm H}$, would thus be $\sim$ 0.006 pc, or about 1300 AU.
Because $n_{\rm H}T$ and $n_{\rm H}$ depend on the assumed $T$, we have listed values for several temperatures consistent with the limit from $b$(\ion{Na}{1}) in Table~\ref{tab:phys}.

Estimates for the electron density in the IV gas may be derived both from considerations of photoionization equilibrium and from \ion{C}{2} excitation equilibrium.
If photoionization equilibrium is assumed for carbon [with $\Gamma$/$\alpha$ = 24 for the WJ1 interstellar radiation field and $T$ = 100 K (P\'{e}quignot \& Aldrovandi 1986) and with $N$(\ion{C}{2}) as estimated above], then $N$(\ion{C}{1})/$N$(\ion{C}{2}) = $n_e$ ($\Gamma$/$\alpha$)$^{-1}$ yields $n_e$(\ion{C}{1}) $\sim$ 6.2$\pm$2.4 cm$^{-3}$.
Analysis of the \ion{C}{2} fine-structure excitation (Fitzpatrick \& Spitzer 1997; Welty et al. 1999) for $T$ = 100 K, however, yields $n_e$(\ion{C}{2}) $\sim$ 0.85$^{+0.52}_{-0.46}$ cm$^{-3}$ --- about a factor of 7 smaller.
While the uncertainties in both $n_e$ estimates are dominated by the almost 50\% uncertainty in $N$(\ion{C}{2}), the ratio of the two estimates is not very sensitive to $N$(\ion{C}{2}).

Similar discrepancies between $n_e$(\ion{C}{1}) and $n_e$(\ion{C}{2}) have been noted for other lines of sight (Fitzpatrick \& Spitzer 1997; Welty et al. 1999).
While \ion{C}{1} might be concentrated in a smaller, denser region than \ion{C}{2}, $n_e$(\ion{C}{1}) for that smaller region would be even larger than the average value derived above, and would imply a fractional ionization $n_e$/$n_{\rm H}$ greater than 25\%.
The $n_e$(\ion{C}{1}) would be reduced, however, if the radiation field were weaker or if $T$ were lower than assumed above.
Alternatively, the large $n_e$ (relative to the assumed WJ1 radiation field) suggests that enough negatively charged large molecules (e.g., PAH's) could be present for charge exchange reactions (X$^+$ + LM$^-$ $->$ X$^0$ + LM$^0$) to significantly enhance the column densities of the trace neutral species --- in which case $n_e$(\ion{C}{1}) would be too high (Lepp \& Dalgarno 1988; Lepp et al. 1988; Welty \& Hobbs 2001).
Even if that is the case, however, $n_e$(\ion{C}{2}) still yields $n_e$/$n_{\rm H}$ $\sim$ 0.034 --- more than a factor 240 greater than the value 1.4 $\times$ 10$^{-4}$ that would be due to photoionization of carbon (usually assumed to be the primary source of electrons).
Hydrogen therefore appears to be partially ($\ga$ 3.4\%) ionized in the $-$38 km s$^{-1}$ cloud.
It is not clear how such a high fractional ionization could be maintained.

\section{Discussion}
\label{sec-disc}

The Leiden-Dwingeloo \ion{H}{1} maps (Hartmann \& Burton 1997) suggest that 
the line of sight to HD 219188 lies near the inner edge of a shell of 
IV gas, seen over the range $-30$ to $-$60 km s$^{-1}$ ($v_{\sun}$).
Despite its possible association with an expanding and/or compressed shell, the cloud's fairly low $T$,
fairly high $N$(\ion{Na}{1})/$N$(\ion{Ca}{2}) ratio, moderate pressure, and stable
velocity suggest that it is relatively quiescent.
The variations in absorption may instead be due to the relatively high proper
motion of HD 219188 --- about 13.4 mas/year (ESA 1997) --- which 
corresponds to a transverse velocity of about 180 km~s$^{-1}$, or about 
38 AU per year, at 2800 pc.
The steadily increasing column densities of \ion{Na}{1} and \ion{Ca}{2} and 
the high fractional ionization suggest that the 
line of sight is moving into/through the partially ionized edge of the
IV cloud due to the motion of the background star.

Unfortunately, the distance to this IV cloud is not well constrained.
A lower limit of 150 pc is provided by the lack of IV absorption 
toward $\beta$ Psc ($<$2\arcdeg away, but toward detectable IV \ion{H}{1} 
emission); the upper limit (2800 pc) is set by HD 219188 itself.
The length scale in the IV cloud probed by the stellar proper motion is 
thus between 2 and 38 AU/year --- considerably smaller than the 10$^2$--10$^4$
AU scales sampled by the binary star studies.
The cloud distance (and thus the length scale) could be better constrained by obtaining \ion{Na}{1} spectra of 
other stars in this region (at various distances).

While we do not have a spectrum of \ion{Na}{1} from 1994--1995, when the UV spectra were obtained, estimates based on the \ion{C}{1} and \ion{Ca}{2} observed then suggest that $N$(\ion{Na}{1}) could have been $\sim$ 2 $\times$ 10$^{11}$ cm$^{-2}$.
The ratio $N$(\ion{Na}{1})/$N$(H) $\sim$ 4.2 $\times$ 10$^{-7}$ would thus have been much higher than is typically observed for diffuse clouds with $N$(H) $\ga$ 10$^{19}$ cm$^{-2}$ (Welty \& Hobbs 2001) --- presumably due to the high fractional ionization.
Furthermore, if Na is depleted by a factor 4 in the IV gas (Welty \& Hobbs 2001), then \ion{Na}{1} would have been the dominant form of Na.

If such high $N$(\ion{Na}{1})/$N$(H) ratios and fractional ionizations are representative of other sightlines where differences in \ion{Na}{1} have been observed over small transverse scales, then the corresponding differences in $N$(H) --- and the resulting inferred pressures and densities --- would not be as large as previously estimated. 
Lauroesch et al. (2000) have estimated $n_e$/$n_{\rm H}$ $\sim$ 0.008 for the variable component at $v$ $\sim$ 9 km s$^{-1}$ toward HD 32040, and $N$(\ion{Na}{1})/$N$(H) is likely high [for $N$(H) $\sim$ 6 $\times$ 10$^{18}$ cm$^{-2}$] in the $-$8.6 km s$^{-1}$ component toward $\mu^1$ Cru (Meyer \& Blades 1996).
The high $N$(\ion{Na}{1})/$N$(H) ratios found for a sample of high- and intermediate-velocity clouds with $N$(H) $\la$ 10$^{19}$ cm$^{-2}$ (Wakker \& Mathis 2000) may also generally be due to partial ionization of the hydrogen. 
These lower $N$(H) clouds may thus represent a different regime from the more familiar diffuse neutral clouds with $N$(H) $\ga$ 10$^{19}$ cm$^{-2}$.

Lauroesch et al. (1998) have conjectured that some of the small-scale
structure observed in \ion{Na}{1} may be due to fluctuations in ionization
[rather than in overall $N$(H)] for cases where corresponding
differences are not seen for \ion{Zn}{2} [which should track $N$(H)].
The low overall column densities, moderate local densities, and high fractional ionization found for the IV cloud toward HD 219188 suggest that some of the small-scale structure seen in \ion{Na}{1} and \ion{K}{1} may be due to partially ionized, relatively low-density cloud edges and/or fragments --- rather than dense, neutral clumps.
Further monitoring of the HD 219188 line of sight over the next several years --- particularly in the UV with {\it HST}/STIS --- should provide interesting information on the variations of density, temperature, and depletions with depth in the IV cloud.

\acknowledgements

We thank D. Willmarth (KPNO), S. Ryan (AAO), and M. Kuerster (ESO) for their help in setting up the spectrographs, Lew Hobbs for pointing out the large proper motion of HD 219188, and Bart Wakker for providing a map of the 21 cm emission (and individual spectra) near HD 219188.
DEW acknowledges support from NASA LTSA grant NAGW-4445.

\begin{deluxetable}{llcccccccc}
\tabletypesize{\scriptsize}
\tablecolumns{10}
\tablenum{1}
\tablecaption{Optical Data:  $v$ = $-$38 km s$^{-1}$ Component\tablenotemark{a}\label{tab:opt}}
\tablewidth{0pt}

\tablehead{
\multicolumn{1}{l}{Date}&
\multicolumn{1}{c}{Observatory}&
\multicolumn{1}{c}{Resolution}&
\multicolumn{3}{c}{-~-~-~-~-~-~Na I~-~-~-~-~-~-}&
\multicolumn{3}{c}{-~-~-~-~-~-~Ca II~-~-~-~-~-~-}&
\multicolumn{1}{c}{$N$(Na I)/$N$(Ca II)}\\
\multicolumn{1}{c}{ }&
\multicolumn{1}{c}{(Instrument)}&
\multicolumn{1}{c}{ }&
\multicolumn{1}{c}{$N_{10}$\tablenotemark{b}}&
\multicolumn{1}{c}{$b$}&
\multicolumn{1}{c}{$v$}&
\multicolumn{1}{c}{$N_{10}$\tablenotemark{b}}&
\multicolumn{1}{c}{$b$}&
\multicolumn{1}{c}{$v$}&
\multicolumn{1}{c}{ }\\
\multicolumn{1}{c}{ }&
\multicolumn{1}{c}{ }&
\multicolumn{1}{c}{(km s$^{-1}$)}&
\multicolumn{1}{c}{(cm$^{-2}$)}&
\multicolumn{1}{c}{(km s$^{-1}$)}&
\multicolumn{1}{c}{(km s$^{-1}$)}&
\multicolumn{1}{c}{(cm$^{-2}$)}&
\multicolumn{1}{c}{(km s$^{-1}$)}&
\multicolumn{1}{c}{(km s$^{-1}$)}&
\multicolumn{1}{c}{ }}
\startdata
1980.75\tablenotemark{c}  & McDonald (2.7m)& 5.9/5.4\tablenotemark{c}  & 
                            $<$3.0 & \nodata & \nodata &
                            4$\pm$1 & (1.5) & ($-$38.6) & $<$0.8 \\
1991.56\tablenotemark{d} & ESO (CAT/CES)& 4.4 & 
                      4$\pm$3 & 1.5$\pm$0.9 & $-$37.9$\pm$1.1 &
                      2$\pm$1 & (1.5) & ($-$38.6) & 2.0$\pm$1.8 \\
1995.82 & KPNO (coud\'{e} feed)& 1.40 & \nodata & \nodata & \nodata &
                        3.1$\pm$0.3 & (1.5) & $-$38.6$\pm$0.2 & \nodata \\
1997.77 & KPNO (coud\'{e} feed)& 1.35 & 27$\pm$1 & 0.59$\pm$0.06 & $-$38.3$\pm$0.1 &
                        4.9$\pm$1.0 & (1.5) & $-$38.9$\pm$0.3 & 5.5$\pm$1.1 \\
1998.68 & AAO (AAT/UCLES) & 5.0 & 34$\pm$2 & (0.6) & $-$38.0$\pm$0.1 &
                      \nodata & \nodata & \nodata & \nodata \\
1999.42 & KPNO (coud\'{e} feed)& 1.50 & 39$\pm$4 & (0.6) & $-$38.1$\pm$0.1 &
                        \nodata & \nodata & \nodata & \nodata \\
1999.98 & ESO (3.6m/CES)& 1.20 & 52$\pm$2 & 0.55$\pm$0.05 & $-$38.2$\pm$0.1 &
                       \nodata & \nodata & \nodata & \nodata \\
2000.46 & KPNO (coud\'{e} feed)& 1.50 & 60$\pm$5 & 0.56$\pm$0.09 & $-$38.3$\pm$0.1 &
                        10.7$\pm$1.2 & (0.6) & $-$38.3$\pm$0.1 & 5.6$\pm$0.8 \\
\enddata
\tablenotetext{a}{Uncertainties are 1$\sigma$; values in parentheses were fixed in the fits.}
\tablenotetext{b}{Column density in units of 10$^{10}$ cm$^{-2}$.}
\tablenotetext{c}{Albert 1983; FWHM for Na I/Ca II; $N$(Ca II) from new fit to generated spectrum.}
\tablenotetext{d}{Sembach et al. 1993; shifted to align low-velocity absorption; $N$(Ca II) from new fit to generated spectrum.}

\end{deluxetable}

\begin{deluxetable}{lccc}
\tablecolumns{4}
\tablenum{2}
\tablecaption{UV Data:  $v$ = $-$38 km s$^{-1}$ Component\tablenotemark{a}\label{tab:uv}}
\tablewidth{0pt}

\tablehead{
\multicolumn{1}{l}{Ion}&
\multicolumn{1}{c}{Line}&
\multicolumn{1}{c}{$N$(X)}&
\multicolumn{1}{c}{[X/S]\tablenotemark{b}}\\
\multicolumn{1}{c}{ }&
\multicolumn{1}{c}{(\AA)}&
\multicolumn{1}{c}{(cm$^{-2}$)}&
\multicolumn{1}{c}{ }}
\startdata
C I    & 1560 & 10.7$\pm$1.0e12 & \nodata \\
C I*   & 1560 &  2.3$\pm$0.3e12 & \nodata \\
C I**  & 1561 &  0.8$\pm$0.3e12 & \nodata \\
C II   & 1334 &  (4.9$\pm$2.1e13)\tablenotemark{c} & ($-$0.4)\tablenotemark{c} \\
C II*  & 1335 &  5$\pm$1e12 & \nodata \\
O I    & 1302 &  1.1$\pm$0.4e14 & $-$0.5 \\
Si II  & 1304 &  9$\pm$1e12 & $-$0.3 \\
%P II   & 1301 & $<$2.6e12 & $<$+1.2 \\
S II   & 1253 &  9$\pm$2e12 & \nodata \\
%Cr II  & 2056 & $<$3.5e11 & $<$+0.2 \\
%Mn II  & 2606 & $<$8.6e10 & $<$$-$0.3 \\
Fe II  & 2374, 2600 & 4.0$\pm$0.4e12 & $-$0.6 \\
%Zn II  & 2026 & $<$5.0e10 & $<$+0.4 \\
\enddata
\tablenotetext{a}{Uncertainties are 1$\sigma$; limits are 2$\sigma$.}
\tablenotetext{b}{[X/S] = log\{$N$(X II)/$N$(S II)\} $-$ log\{(A$_{\rm X}$/A$_{\rm S}$)$_\odot$\}, using the solar system meteoritic abundances of Anders \& Grevesse 1989 and Grevesse \& Noels 1993.}
\tablenotetext{c}{Estimated assuming typical interstellar [C/S] = $-$0.4$\pm$0.1 dex (for C I + C II).}

\end{deluxetable}

\begin{deluxetable}{lccc}
\tablecolumns{4}
\tablenum{3}
\tablecaption{Inferred Properties:  $v$ = $-$38 km s$^{-1}$ Component\label{tab:phys}}
\tablewidth{0pt}

\tablehead{
\multicolumn{1}{l}{Quantity}&
\multicolumn{3}{c}{Value}}
\startdata
$N$(H) (cm$^{-2}$)    & \multicolumn{3}{c}{4.8 $\times$ 10$^{17}$} \\
$T$   (K)  & \multicolumn{3}{c}{$<$490}      \\
\tableline
     & 25 K & 100 K & 400 K \\
\tableline
log($n_{\rm H}T$) (cm$^{-3}$K) & 3.4$\pm$0.1 & 3.4$\pm$0.1 & 3.8$\pm$0.1 \\
$n_{\rm H}$ (cm$^{-3}$) & 100 & 25 & 15 \\
thickness (AU) & 325 & 1300 & 2200 \\
$n_e$(C I)\tablenotemark{a} (cm$^{-3}$) & 2.6 & 6.2 & 14.7 \\
$n_e$(C II)\tablenotemark{b} (cm$^{-3}$) & \nodata & 0.85 & 0.79 \\
$n_e$/$n_{\rm H}$\tablenotemark{c} & \nodata & 0.034 & 0.053 \\
\enddata
\tablenotetext{a}{Assuming photoionization equilibrium for carbon; WJ1 radiation field.}
\tablenotetext{b}{From C II fine-structure excitation equilibrium; no valid value for $T$ = 25 K.}
\tablenotetext{c}{Using $n_e$ from analysis of C II fine-structure.}
\end{deluxetable}

\clearpage 

\end{document}